\documentclass[12pt]{article}
\usepackage{a41}
\usepackage{epsfig}

\newcommand{\nn}{\nonumber}
\newcommand{\ed}{\end{document}}
\newcommand{\be}{\begin{equation}}
\newcommand{\ee}{\end{equation}}
\newcommand{\ba}{\begin{eqnarray}}
\newcommand{\ea}{\end{eqnarray}}
\newcommand{\baz}{\begin{eqnarray*}}
\newcommand{\eaz}{\end{eqnarray*}}
\newcommand{\bb}{\begin{thebibliography}}
\newcommand{\eb}{\end{thebibliography}}
\newcommand{\ct}[1]{${\cite{#1}}$}

\begin{document}

\sloppy
\thispagestyle{empty}

\vspace{1cm}

\mbox{}

\vspace{5cm}

\begin{center}

{\large\bf  Unusual Properties of the Central Production \\
of Glueballs  and Instantons}\\

\vspace*{.5cm}

{\bf N.I.~Kochelev\footnote{kochelev@thsun1.jinr.ru}}\\

{\it Bogoliubov Laboratory of Theoretical Physics,\\
Joint Institute  for Nuclear Research, 141980  Dubna,\\
Moscow Region, Russia}
\end{center}
\vspace*{1cm}
\begin{abstract}

It is shown that  instantons provide a natural mechanism to explain
an unusual azimuthal  dependence of the production 
of the even-parity glueball
candidates in central pp collision. A different 
azimuthal   dependence
for instanton-induced production of the odd-parity glueballs is
predicted.
\end{abstract}
\newpage

Recently,  very interesting experimental data on
central production of glueball candidates  have been published
by  WA91 and WA102 Collaborations \ct{WA91},
\ct{WA102}. These data have shown an unusual
strong  dependence of the cross section for production
of  glueball candidates on
\begin{equation}
dP_t=|p_{1t}^\prime-p_{2t}^\prime |,
\label{trans}
\end{equation}
where $p^\prime_{i t}$, $i=1,2$, are the transverse momentum of
the final  protons.
 The experimental data are specified by a  significant enhancement
  at small values of $dP_t$  of the
production of $f_0(1500), f_J(1710), f_2(1900) $, which are
the candidates for even-parity glueball states \ct{lat},
in comparison with the
known $q\bar q $ mesons.
This experimental observation  opened the door for investigation
of the properties of glueballs in  diffractive processes \ct{closekirk}.
A possible dynamical explanation of this feature, based  on the broken
scale invariance has been suggested in  ref.\ct{ellis}.
However the final result of ref.\ct{ellis} shows only a weak dependence of the
cross section for glueball production at the relative
azimuthal angle $\Phi$ between transverse momenta of the final protons.
It turns out that this weak dependence  is  related
just  with the
kinematics of the double-diffractive production of mesons and is not
connected with some specific  properties of glueballs.

In this letter, we suggest a new mechanism for central glueball
production, based on the instanton structure of the QCD vacuum.

The instantons describe the tunneling between different
gauge-rotated classical vacua in QCD and reflect the nonabelity of theory
of strong interactions. The instanton--induced quark-quark
t'Hooft \ct{thooft} interaction  plays a very important role in
chiral symmetry breaking and in the appearance of the masses of constituent
quarks and hadrons (see reviews \ct{shuryak}, \ct{diakonov}).

It is very important that besides famous quark-quark interaction,
the existence of the instantons in QCD vacuum should lead  to 
specific quark-qluon \ct{koch} and gluon-gluon interactions.

Just  the instanton-induced  gluon-gluon
interaction can give an important contribution to the 
central production of glueballs
in  the $pp$ collision due to the diagram  shown in Fig.1.
It should be mentioned that the possibility of a large instanton
contribution to  the glueball
production  comes from the non-perturbative origin of the instanton
field which gives the factor $1/\alpha_s$ into  the production cross section
for  each gluon that is connected with the instanton.
\begin{figure}[htb]
\centering
\epsfig{file=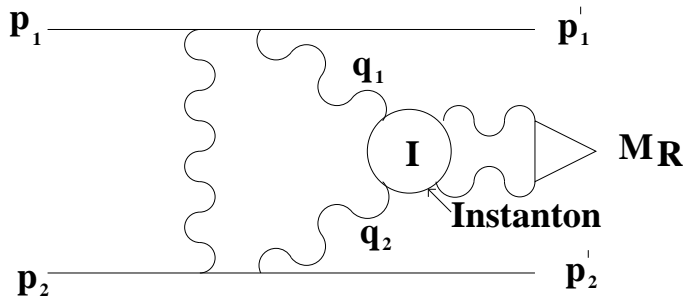,width=9cm}
\vskip 0.5cm
\caption{\it The  contribution to the  central glueball production
 induced by instantons.}
\end{figure}

Let us estimate the $\Phi$ dependence of the instanton-induced
glueball double-diffractive production.
The wave function of gluons incoming onto the instanton
(see Fig.1) is given by the Fourier transform of the instanton
field
\begin{equation}
A_\mu^a(q)=\frac{16\pi^2\rho^2}{g_s}\frac{\bar{\eta}_{a\mu\nu}q_\nu}{q^4}
(1-\frac{1}{2}K_2(\rho|q|)\rho^2q^2)\equiv\bar\eta_{a\mu\nu}q_\nu f(q^2),
\label{gluon}
\end{equation}
where the t'Hooft's symbol $\bar\eta_{a\mu\nu} $ 
in the Minkowsky space is
\begin{equation}
\bar\eta_{aij}=\epsilon_{aij}, {\  } \bar\eta_{a0i}=i\delta_{ai},
\label{eta}
\end{equation}
 and $\bar\eta_{a\mu\nu} $  is antisymmetric under the interchange of $\mu$
and $\nu$.

At high energies the matrix element  of the reaction presented in Fig.1
is given by the formula
\begin{equation}
M\approx\bar\eta_{a\mu\nu}\bar\eta_{a\mu^\prime\nu^\prime}
q_{1\nu}q_{2\nu^\prime}\tilde{p}_{1\mu}\tilde{p}_{2\mu^\prime}
F({q_1}^2,{q_2}^2),
\label{matrix}
\end{equation}
where $q_i=p_i-p_i^\prime$, $\tilde{p}_i=p_i+p_i^\prime$, and
$F({q_1}^2,{q_2}^2)$ is a form factor which contributes negligibly
  to the $\Phi$ dependence and therefore
we will neglect it.

To obtain formula (\ref{matrix}), the leading-gluon
model for the pomeron has been used (see a discussion
in \ct{ellis}). In this model the meson production (Fig.1) is determined by
the fusion of the two leading gluons from  different pomerons
whereas the remained gluons from the pomerons  provide  only 
colourlessness.

The contraction of the t'Hooft's symbols Eq.(\ref{matrix}) is
\begin{equation}
\bar\eta_{a\mu\nu}\bar\eta_{a\mu^\prime\nu^\prime}=
M^{even}_{\nu\mu\nu^\prime\mu^\prime}+M^{odd}_{\nu\mu\nu^\prime\mu^\prime}
\label{parity}
\end{equation}
where
\begin{eqnarray}
M^{even}_{\nu\mu\nu^\prime\mu^\prime}
=g_{\nu\nu^\prime}g_{\mu\mu^\prime}
-g_{\nu\mu^\prime}g_{\nu^\prime\mu},
\label{even}
\end{eqnarray}
and
\begin{eqnarray}
M^{odd}_{\nu\mu\nu^\prime\mu^\prime}=
\epsilon_{\nu\nu^\prime\mu\mu^\prime}.
\label{odd}
\end{eqnarray}
It is easy to check that two terms in (\ref{parity})
contribute to the production of mesons with opposite
parities. This follows from the fact that one can rewrite the even
part of the amplitude (\ref{parity}), (\ref{even}) through
the even-parity product of the field strengths of  incoming gluons
$G^a_{\mu\nu}G^a_{\mu^\prime\nu^\prime}$,
 whereas the odd-parity part (\ref{odd}) can be represented by the formula
$G^a_{\mu\nu}\widetilde{G}^a_{\mu^\prime\nu^\prime}$,
where $\widetilde{G}^a_{\mu^\prime\nu^\prime}=
\epsilon_{\mu^\prime\nu^\prime\rho\tau}G^a_{\rho\tau}/2$.
Therefore in the production of the even- (odd-) parity glueballs
only the first (second) term in (\ref{parity}) can contribute.

In the center of mass system of  initial protons the momenta
of  initial and final particles are
\begin{eqnarray}
p_1&\approx&(P,0,P), {\ }{\ }{\ } {\ } p_2\approx(P,0,-P)\nn\\
{p_1}^\prime&\approx&(x_1P,\vec{p}_{1t},x_1P),
{\ }{\ }{p_2}^\prime\approx(x_2P,\vec{p}_{2t},-x_2P)
\label{kinematic}
\end{eqnarray}
By using this formula and (\ref{even}), (\ref{odd}) one can obtain
the following $\Phi$ dependence of the even- and odd-parity
glueball production cross-section induced by instantons:
\begin{equation}
\sigma_{even}\sim |M|^2\sim (\vec{p}_{1t}.\vec{p}_{2t})^2\sim cos^2\Phi,
\label{cseven}
\end{equation}
\begin{equation}
\sigma_{odd}\sim |M|^2\sim |\vec{p}_{1t}\times\vec{p}_{2t}|^2\sim sin^2\Phi,
\label{csodd}
\end{equation}

To predict the total $\Phi$ dependence of the glueball production, one
should take into account the kinematic restrictions for the
meson production \ct{ellis}. For the
symmetric kinematics $x_1=x_2=x_P-1$, $t_1\approx t_2$, by using the pomeron
flux factor
\begin{equation}
f_P(x_P)\propto x_P^{1-2\alpha(t)},
\nn
\end{equation}
 where $\alpha(t)=1+\epsilon+\alpha^\prime t$ is the pomeron
trajectory with $\epsilon=0.085$ and $\alpha^\prime=0.25 GeV^2$,
one can obtain from the double-diffractive cross section formula
\footnote{ The kinematic relation $x_P=(M_R^2+2|t|(1-cos\Phi))^{1/2}
/\sqrt{s}$ \ct{ellis}, where $M_R$ is the mass of the glueball was
used  to obtain (\ref{final1}), (\ref{final2}).}
\begin{equation}
\frac{d\sigma}{d\Phi}\approx f^2_P(x_P)\frac{d\hat{\sigma}}{d\Phi}
\nn
\end{equation}
the following $\Phi$ dependence of even- and odd-parity glueball
production
\begin{equation}
\frac{d\sigma^{even}}{d\Phi}\sim\left(\frac{s}{M_R^2+2|t|(1-cos\Phi)}
\right)^{1+2\epsilon+2\alpha^\prime t}cos^2\Phi
\label{final1}
\end{equation}
and
\begin{equation}
\frac{d\sigma^{odd}}{d\Phi}\sim\left(\frac{s}{M_R^2+2|t|(1-cos\Phi)}
\right)^{1+2\epsilon+2\alpha^\prime t}sin^2\Phi.
\label{final2}
\end{equation}
The distribution  (\ref{final1}) for the lightest scalar glueball
candidate $f_{0}(1500)$
 production for\\ $t=-0.1 GeV^2$ in  WA91 and WA102 experiments
is shown in  Fig.2. This distribution  significantly differs from the
distribution  predicted within  model \ct{ellis}.
The difference is connected with a
much more larger enhancement of the  even-parity
glueball production
at small values of  $\Phi$ and its suppression at
$\Phi\approx\pi/2$.
This dependence is in qualitative agreement with experimental data
\ct{WA91}, \ct{WA102}.

\begin{figure}[htb]
\centering
\begin{minipage}[c]{7.5cm}
\vskip -1.1cm
\centering
\epsfig{file=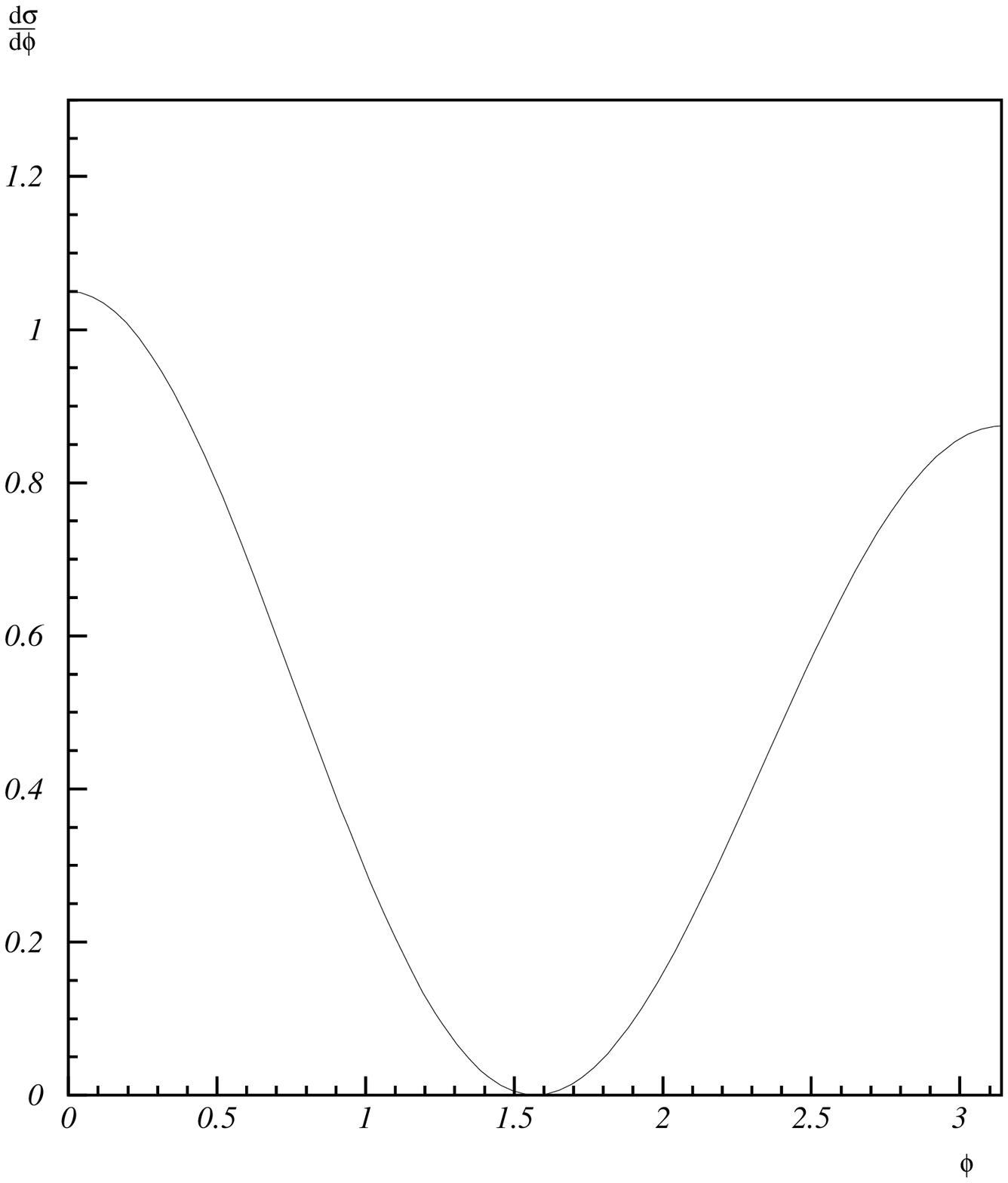,width=7.5cm}
\caption{\it $\Phi$-dependence of the instanton contribution to
 $f_0(1500)$ even-parity meson central production  at $
 t=-0.1GeV^2$.}
\end{minipage}
\hspace*{11mm}
\begin{minipage}[c]{7.5cm}
\centering
\epsfig{file=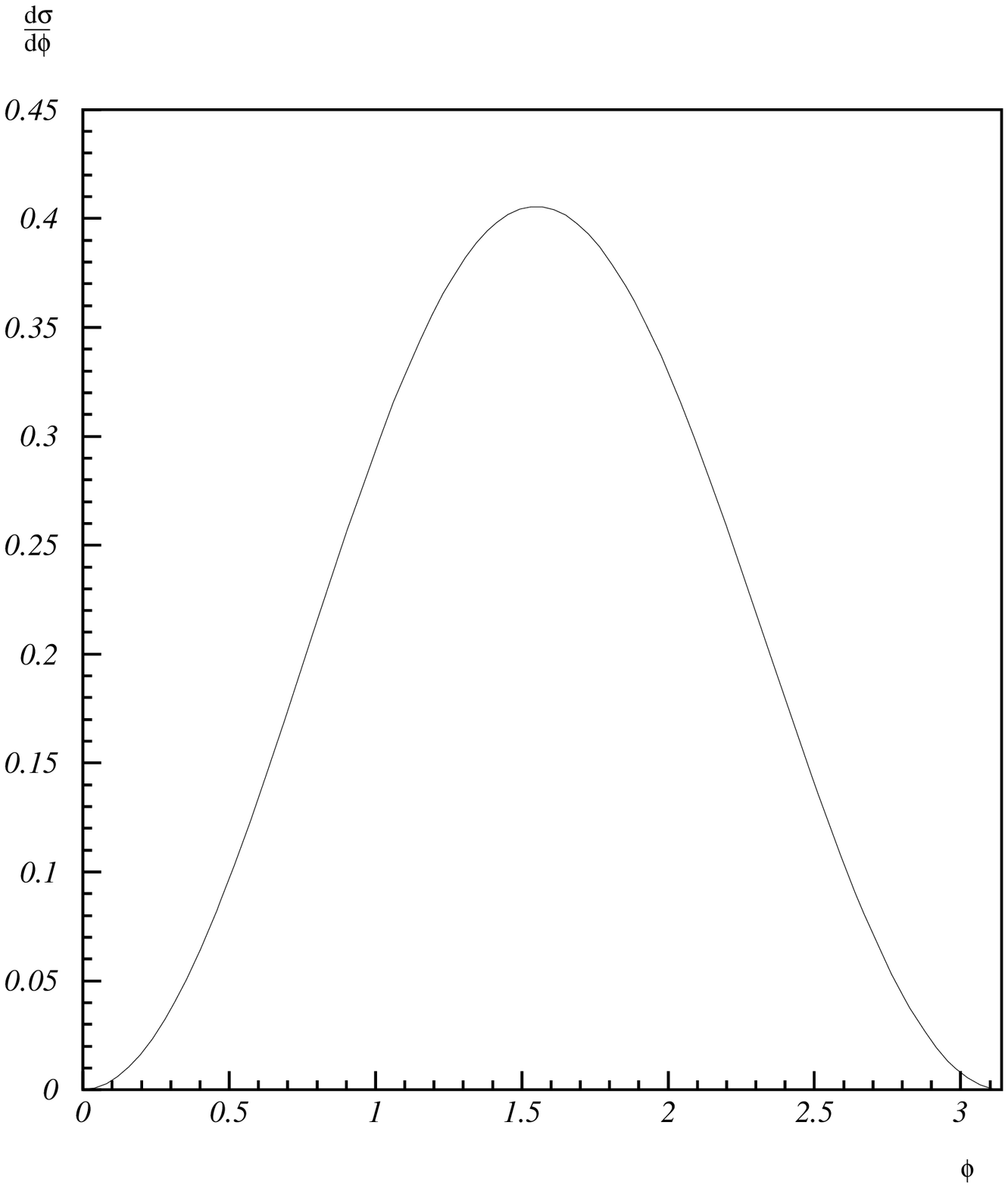,width=7.5cm}
\caption{\it $\Phi$- dependence of the instanton contribution to
the  central production of  odd-parity glueball with
 mass $M_R=2250 MeV$ at $t=-0.1 GeV^2$.
\protect\rule[-\baselineskip]
 {0pt}{2\baselineskip} }
\end{minipage}
\end{figure}

For the production of the  odd-parity  glueballs there should be an opposite
situation. Indeed, their cross section should be small at small values of
 $\Phi$ and significantly larger at  $\Phi\approx\pi/2$.
The corresponding distribution for the production of the
odd-parity   lightest glueball candidate $0^{-+}$ with the mass 
$M_R(0^{-+})\approx 1.5M_R(0^{++})\approx 2250 MeV$
\ct{lat} is presented in Fig.3.

Recently the experimental data on the production of pseudoscalar neutral mesons
$\pi^0$,
$\eta$, $\eta^\prime$ in the central pp collision have been
published \ct{WA102mesons}. It was discovered that the production
mechanism for $\eta$ and $\eta^\prime$  differs from the
mechanism of the $\pi^0$ production. Indeed, it was found  that  the cross
section for the $\eta$ and $\eta^\prime $ production are greatest when
the azimuthal angle $\Phi$ between
$p_t$ of  two final protons  is $\pi/2$.
From our point of view,  due to the large value of matrix 
elements~\ct{shifman} 
\begin{eqnarray} 
<0|\frac{3\alpha_s}{4\pi}G^a_{\mu\nu}\widetilde{G}^a_{\mu\nu}|\eta>=
\sqrt{\frac{3}{2}}f_\pi m_\eta^2,\nn\\
<0|\frac{3\alpha_s}{4\pi}G^a_{\mu\nu}\widetilde{G}^a_{\mu\nu}|\eta^\prime>=
\sqrt{3}f_\pi m^2_{\eta^\prime}
\label{shifman}
\end{eqnarray}
the diagram in Fig.1 can  significantly contribute also to
$\eta$ and $\eta^\prime$ production, which leads to the enhancement of
production at $\Phi\approx\pi/2$.

In summary, the  new mechanism for  central production of glueballs 
is suggested. It is based on the instanton picture of the QCD
vacuum and provide a natural explanation of the unusual kinematical
dependence of  cross sections of production of glueball candidates.
This observation gives a nice opportunity to investigate the
properties of the complicated QCD vacuum in the  double-diffractive
production of mesons.

 The author is very grateful to  A.E.Dorokhov, S.B.~Gerasimov, A.~Kirk
and V.~Vento for useful discussions.


\begin{thebibliography}{99}
\bibitem{WA91} D.Barberis et al., WA91 Collaboration, {\it Phys. Lett.}
{\bf B388} (1996) 853.
\bibitem{WA102} D.Barberis et al., WA102 Collaboration, {\it Phys. Lett.}
{\bf B397} (1997) 339;\\
A.Kirk et al., WA102 Collaboration, hep-ph/9810221.
\bibitem{lat} G.Bali et al. (UKQCD), {\it Phys. Lett.} {\bf B309} (1993) 378;\\
D.Weingarten, hep-lat/9608070;\\
J.Sexton et al., {\it Phys. Rev. Lett.} {\bf 75} (1995) 4563.
\bibitem{closekirk} F.E.Close and A.Kirk, {\it Phys. Lett.} {\bf B397}
(1997) 333.
\bibitem{ellis} J.Ellis and D.Kharzeev, CERN-TH/98-349, RIKEN-BNL preprint,
hep-ph/9811222.
\bibitem{thooft}
 't Hooft {\it Phys. Rev.} {\bf D14} (1976) 3432.
\bibitem{shuryak} T.Sch\"afer and E.V.Shuryak,
{\it Rev. Mod. Phys.} {\bf 70} (1998) 1323.
\bibitem{diakonov} D.I.Diakonov, hep-ph/9602375.
\bibitem{koch} N.I.Kochelev,
{\it Phys. Lett.} {\bf B426} (1998) 149.
\bibitem{WA102mesons} D.Barberis et al., WA102 Collaboration,
 {\it Phys. Lett.} {\bf B427} (1998) 398.
\bibitem{shifman} V.A.Novikov, M.A.Shifman, A.I.Vainshtein, A.I.Zakharov,
{\it Nucl. Phys.} {\bf B165} (1980) 55;\\
M.A. Shifman, {\it Phys. Rep.} {\bf 209} (1991) 341.
\end{thebibliography}
\end{document}